# Observation of topological states residing at step edges of WTe$_2$

Lang Peng[1§], Yuan Yuan[1§], Gang Li[2,3#], Xing Yang[1], Jing-Jing Xian[1], Chang-Jiang Yi[4], You-Guo Shi[4], Ying-Shuang Fu[1*]

1. School of Physics and Wuhan National High Magnetic Field Center, Huazhong University of Science and Technology, Wuhan 430074, China
2. School of Physical Science and Technology, ShanghaiTech University, Shanghai 200031, China
3. Institute of Solid State Physics, Vienna University of Technology, A-1040 Vienna, Austria
4. Institute of Physics, Chinese Academy of Sciences, Beijing 100084, China

Email: *yfu@hust.edu.cn  #ligang@shanghaitech.edu.cn

§ These authors contribute equally to this work.

**Topological states emerge at the boundary of solids as a consequence of the nontrivial topology of the bulk. Recently, theory predicts a topological edge state on single layer transition metal dichalcogenides with 1T' structure. However, its existence still lacks experimental proof. Here, we report the direct observations of the topological states at the step edge of WTe$_2$ by spectroscopic-imaging scanning tunneling microscopy. A one-dimensional electronic state residing at the step edge of WTe$_2$ is observed, which exhibits remarkable robustness against edge imperfections. First principles calculations rigorously verify the edge state has a topological origin, and its topological nature is unaffected by the presence of the substrate. Our study supports the existence of topological edge states in 1T'-WTe$_2$, which may envision in-depth study of its topological physics and device applications.**

In the two-dimensional (2D) topological insulators[1,2], the nontrivial edge states (ES) support quantum spin Hall (QSH) effect, where the electrons at the edge of the system possess different spins when propagating along opposite directions. This enables the topological ES immune to scattering from non-magnetic impurities as protected by the time-reversal symmetry. Since the early proposal of QSH effect in graphene with spin-orbit coupling[3] (SOC), a number of other QSH systems that are more readily examined by experiments are predicted[4]. However, only a few of them are confirmed experimentally[5-11]. Recently, a family of QSH insulators based on single-layer transition metal dichalcogenides (TMDs) with 1T' structure is predicted theoretically[12]. Combining the capability of van der Waals (vdW) stacking, the TMD QSH insulators provide the advantage of multiple edge conduction channels, which is highly desirable for practical device applications. This stimulates intensive research interests in identifying the topological phases in single layer TMDs, especially how to tune the crystal structure into the 1T' phase[13-15]. Despite that a few 1T'-TMD single layers and devices have been successfully synthesized[16,17] or fabricated[18,19], conclusive remarks on the topological ES have still not been reached.

$WTe_2$ has the 1T' structure in its natural ground state and exhibits the largest SOC strength among TMDs. These properties render it the most promising candidate to search for the predicted QSH state[12]. Moreover, $WTe_2$ is predicted to possess nontrivial topological phases in its bulk as type-II Weyl fermions[20]. Step edges of the bulk $WTe_2$ offer the interfaces between the $WTe_2$ layer and the vacuum, which are distinct in electronic topology. In this regard, compared to the monolayer $WTe_2$ that is challenging to fabricate, the bulk step edges may be an appealing alternative to examine the potentially hosted QSH state. For this, we study the electronic states of step edges of $WTe_2$ with spectroscopic imaging scanning tunneling microscopy (SI-STM), which is a powerful probe sensitive to the local density of state (LDOS) with high energy and high spatial resolution.

$WTe_2$ has a layered structure with vdW bonding between the layers. Its 1T' structure originates from a lateral distortion of W atoms towards the *b* direction of the 1T structure. This creates zigzag chains along the *a* direction with neighboring chains having different heights (Fig. 1a). The distortion directions of the adjacent $WTe_2$ layers are opposite. Thus, the "top" and "bottom" surfaces are of different type. After cleaving, both surfaces can sometimes coexist that are separated with single-layer high step edges. Figure 1c shows a STM image of such surfaces with a step. Its zoomed-in image clearly resolves (Fig. 1b) the atomic resolution of the Te atoms of the $WTe_2$ (001) surface, which exhibits alternating bright and dark chain structures. Standing wave patterns in the vicinity of the step edges and defects seen in Fig. 1c propagate along the chain direction, reflecting the anisotropic character of the band structure in $WTe_2$. The step height is measured as ~0.72 nm, i.e. single layer high $WTe_2$, demonstrating the upper and lower terraces are two different surfaces. Such assignment is augmented by the appearance of defects, which mark the type of surfaces via symmetry. Figure 1d and e show two identical kinds of defects at both surfaces. They point to opposite directions, clearly revealing the two surfaces are different.

To unravel its electronic structure, we acquire tunneling spectra (Fig. 2a, left), which is proportional to the LDOS, along a line perpendicularly across the step edge in the *a*-direction (Fig. 1c, black line and Fig. 2a, right). It is seen that the spectra at the inner terraces are spatially homogenous and are identical on both surfaces. Upon approaching the step edge, the tunneling conductance enhances drastically than that of the inner terrace between ~50 meV and ~130 meV. This indicates the existence of prominent ES. To inspect more details, we extract the spectra obtained at the step edge (Fig. 2b, red line) and at the inner terrace (Fig. 2b, black line). Evidently, the tunneling conductance of the inner terrace has finite value for all energies, consisting with $WTe_2$ being a semi-metal. Its spectroscopic shape is captured nicely with our calculated density of states (DOS) with density functional theory (DFT) (Fig. 2c and Supplementary Fig. S1). For the spectroscopy at the step edge, there appear two peaks at ~60 meV and 200 meV, respectively. The detailed shape of the ES varies at different locations of the step edge (Supplementary Fig. S2), indicating its edge configuration is not uniform.

Next, we did real space spectroscopic mapping to a straight step (Fig. 2d) along the *a*-direction to clarify the nature of the ES. Remarkably, the ES intensity enhances between ~50 meV and 120 meV, and gradually depresses outside the energy window. (Spectroscopic mappings at representative energies are shown in Fig. 2e). The ES is localized precisely along the step edge, revealing its origin and also confirming it is a 1D state. A conductance profile across the ES shows it has a lateral spatial extension of ~2.5 nm (Fig. 2f). This is of similar size as the topological ES observed in other systems[7,8,10].

More importantly, scrutiny on an irregular shaped step edge (Fig. 3a) indicates the robustness of the ES. Remarkably, the conductance intensity of the ES exhibits the identical energy dependence (Fig. 3b) and the equal lateral spatial distribution (Fig. 3c) as the straight edge in Fig. 2d. It is noted that the spectroscopic shape of the ES varies at different locations of the irregular step edge (Fig. 3d), resulting in conductance fluctuations of the ES intensity. Nevertheless, the inspected locations entirely exhibit the ES prominently. This demonstrates the emergence of the ES is not specific to a particular step direction, but is related to the nontrivial topology of the bulk.

To confirm this hypothesis, we perform density-functional calculations. We start with a freestanding monolayer 1T'-WTe$_2$ ribbon by using the tight-binding model constructed from the bulk electronic structure. Its edges are terminated equivalently on both sides (Fig. 4a, inert), where no edge potential and structure relaxation are accounted. The ribbon width is infinitely large to exclude hybridizations between the two edges. Both a $n$-field approach[21,22] and a hybrid Wannier charge center[23] (Wilson loop[24]) method confirm the existence of topological ES (Fig. 4a, b and Supplementary Section 3.1). As a consequence of the nontrivial topology of the bulk, the existence of the topological ES should be irrelevant to the specific edge geometry. Indeed, our calculation confirms its existence at both $b$-edge (Supplementary Fig. S4) and differently terminated $a$-edges (Supplementary Fig. S5). It is found that the detailed spectroscopic features of the ES alter with different edge terminations, recalling the experimental observations. (Supplementary Fig.S2).

We then take the edge potential and structure relaxation into account, and calculate a 1T'-WTe$_2$ monolayer ribbon of 62 Å wide with the same edge termination as in Fig. 4a directly with *ab-initio* calculation. As seen in Fig. 4c, the topological ES are clearly present with an increased energy separation compared to Fig. 4a. This calculated result is qualitatively consistent with Ref. 12, despite the details of the band features are different owing to the different edge geometry and exchange potential.

To further evaluate the substrate effect, we consider a monolayer 1T'-WTe$_2$ ribbon of 30 Å width placed on another layer of it with *ab-initio* calculation. The ES on the step edge shown in Fig. 4d not only qualitatively but also quantitatively resembles that of the freestanding monolayer edge (Fig. 4c). Furthermore, the charge distribution of the ES locates on the edge of the upper terrace, which substantiates their origin from the edge (Fig. 4e and 4f) and agrees with the experimental findings in Fig. 2e. This indicates the interlayer coupling does not alter the topological property of the ES due to the weak coupling nature of the vdW stacking. This may have implications to other layer-stacked topological materials, which promises an abundance of unexplored experimental possibilities. Moreover, bulk WTe$_2$ is predicted to be a type-II Weyl semimetal, which features 4 small sections of disconnected topological surface Fermi arcs at its (001) surface[20]. The topological surface Fermi arcs are distinct from the topological ES at the step edge in momentum space, preserving the 1D nature of the topological ES[10].

Our study thus not only provides an experimental proof to the QSH ES in monolayer WTe$_2$ albeit indirectly, but also expands the accessible scope of such nontrivial states to bulk samples. This is of particular importance for situations where bulk properties are desirably introduced into the system. For instance, bulk WTe$_2$ show superconducting phase transition under pressure[25,26]. We have examined theoretically that the topological ES is preserved in the pressure regime where superconductivity of bulk WTe$_2$ coexists (Supplementary Section 4). This renders the step edge of WTe$_2$ a promising system for realizing topological superconductivity[27]. We envision that multiple conduction channels can be created for device applications by forming step edge arrays through the technologically compatible lithographic patterning technique, which is however infeasible for the monolayer, or further growth of nanostripes on WTe$_2$ substrate.

*Note added in proof*: After completion of this manuscript, we became aware of related work[28, 29].

**Methods**

The experiments were performed with a custom-made Unisoku STM (1300) at 4.4 K[30]. WTe$_2$ crystals grown by a solid-state reaction were cleaved *in situ* under ultrahigh vacuum conditions at ~ 77 K. After cleaving, the crystals were transferred quickly to the low temperature STM for subsequent measurements. An electro-chemically etched W wire was used as the STM tip. Prior to measurements, the tip was characterized on a Ag(111) crystal that had been cleaned by several cycles of Ar ion sputtering and annealing. The tunneling spectra were obtained by lock-in detection of the tunneling current with a modulation voltage at 983 Hz feeding into the sample bias. The WTe$_2$ monolayer was calculated with density functional theory. The projector-augmented-wave[31,32] method implemented in the Vienna Ab Initio Simulation Package[33,34] was employed with an energy cutoff of 500 eV. The generalized gradient approximation potential[35] was used in all calculations. The topological ES was calculated by applying the iterative Green's function approach[36] based on the maximally localized Wannier functions[37] obtained through the VASP2WANNIER90[38] interfaces in a non-self-consistent calculation with 9×9×1 k-mesh.

**Acknowledgement:**

The authors thank G. Xu, T. Hanaguri, S.-Q. Shen, Z.-W. Zhu, and X. Liu for helpful discussions. This work is funded by the National Science Foundation of China (Grants No. 11474112, No. 11522431, No. 11474330), the National Key Research and Development Program of China (Grant No. 2016YFA0401003, 2016YFA0300604). G. Li acknowledges the starting grant of ShanghaiTech University.

**Author contributions:**

P.L. and Y.Y. carried out the experiments with the help of X.Y. and J.J.X.. G.L. did the theoretical calculations. Y.G.S. and C.J.Y. grew the $WTe_2$ single crystal. Y.S.F. supervised the project. Y.S.F. and G.L. designed the experiment, analyzed the data and wrote the manuscript.

**Competing financial interests**

The authors declare no competing financial interests.

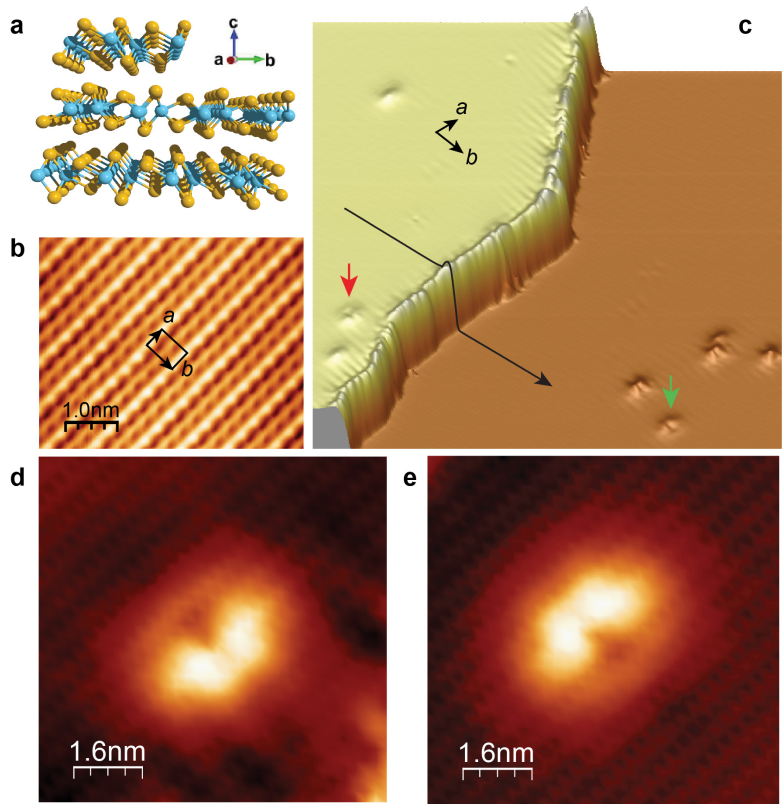

**Figure 1 Topography of WTe$_2$. a**, Schematic illustrating the crystal structure of WTe$_2$ with a step edge. The W and Te atoms are colored with cyan and orange, respectively. **b**, STM image showing the topography of WTe$_2$ with atomic resolution.   Imaging conditions: $V_s$ = 120 mV, $I_t$ = 100 pA. The unit cell of the (001) plane is marked. **c**, Pseudo 3D image of WTe2 showing a step edge. Image size: 60×60 nm. The black line is a sectional line drawn across the step. The bright protrusions on both the upper and lower terraces are crystal defects. The red arrow and green arrow mark two typical defects that are mirror-symmetric to each other. Their zoom-in STM images are shown in **d** and **e**, respectively. The wave like patterns around the defects and steps in **c** are the electron standing waves. Imaging conditions of **c, d** and **e**: $V_s$ = 150 mV, $I_t$ = 200 pA.

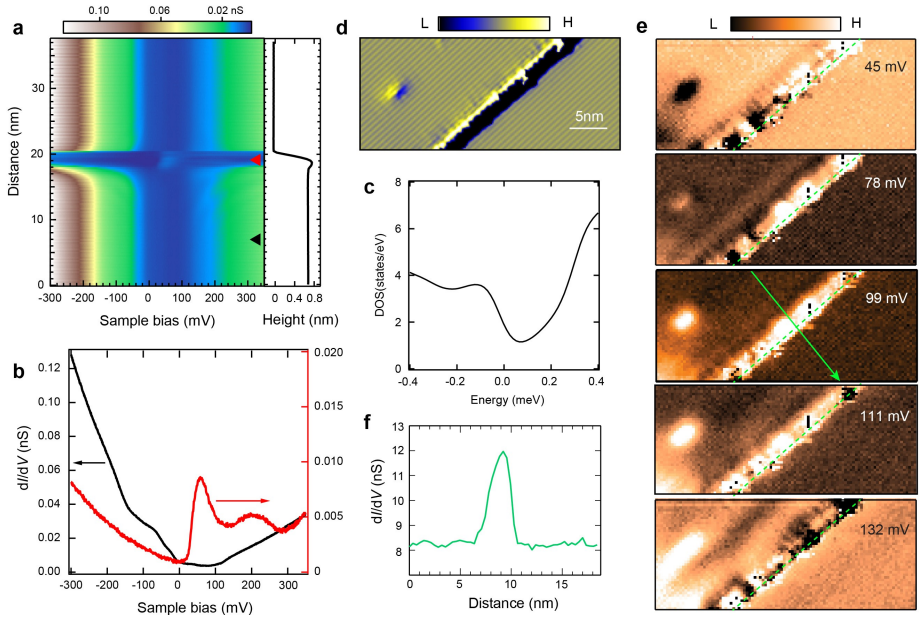

**Figure 2 Spectroscopy of the edge states of WTe$_2$. a,** Two-dimensional conductance plot of tunneling spectra (left) measured along the black line in Fig. 1**c**, whose sectional line is shown in the right, showing the emergence of edge states around the step edge. **b,** Typical tunneling spectra measured at the step edge (red curve) and at a location at the inner terrace (black curve). The spectra were extracted from **a** at locations marked with red and black triangles, respectively. Measurement conditions of **a** and **b**: $V_s$ = 150 mV, $I_t$ = 200 pA and $V_{mod}$ = 3.5 mV$_{rms}$. **c,** Calculated density of states of bulk WTe$_2$ with DFT. **d,** Derivative STM image of a single layer high step edge along *a*-direction. The left terrace is the higher terrace. Measurement conditions: $V_s$ = 180 mV, $I_t$ = 200 pA. **e,** Spectroscopic mapping of the imaged area in **d** at different voltages, showing the spatial distribution of the edge state with energy. Measurement conditions: $V_s$ = 180 mV, $I_t$ = 600 pA, $V_{mod}$ = 3.5 mV$_{rms}$. **(f)** A sectional line profile extracted from **e** along the green line. The green dashed lines in **e** mark the position of the step edge.

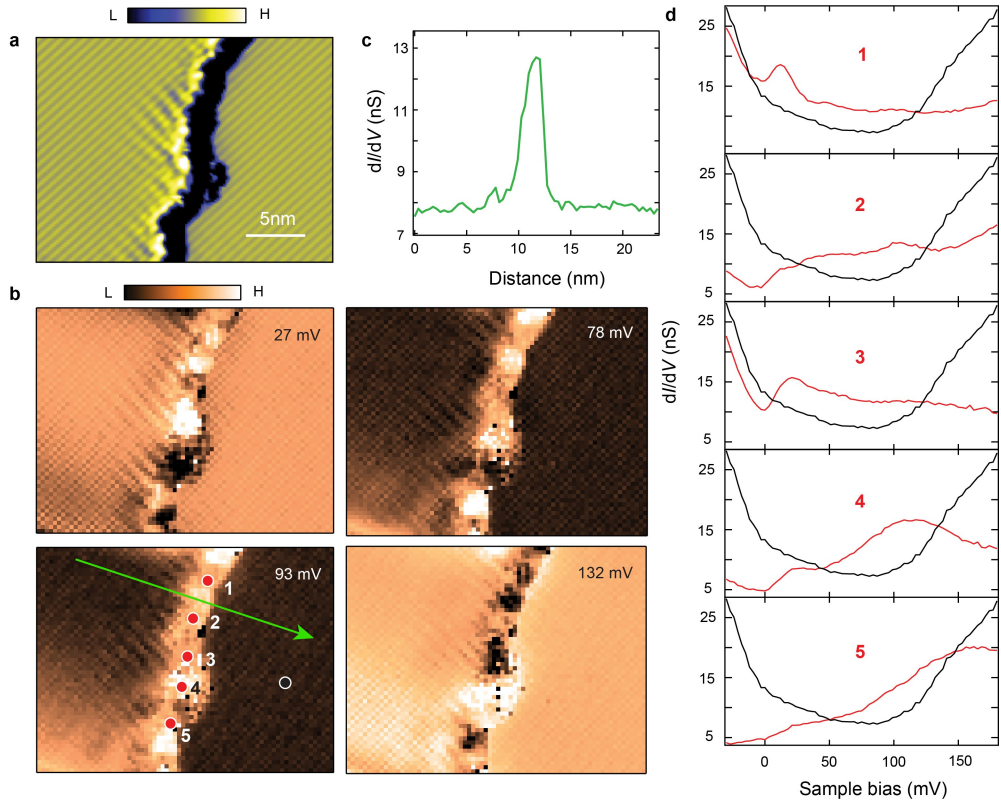

**Figure 3 Edge states of WTe$_2$ residing at an irregular shaped step edge. a**, Derivative STM image of a single layer high step edge with irregular shape. Measurement conditions: $V_s$ = 180 mV, $I_t$ = 200 pA. **b**, Spectroscopic mapping of the imaged area in **a** at different voltages, showing the spatial distribution of the edge state with energy. **c**, A sectional line profile extracted from **b** along the green line. **d**, Tunneling spectra (red curves) at different locations of the step edge (red dots in **b**). The spectroscopy (black curve) of the inner terrace (black dot in **b**) is shown for comparison. Measurement conditions of **b** and **d**: $V_s$ = 180 mV, $I_t$ = 600 pA, $V_{mod}$ = 3.5 mV$_{rms}$.

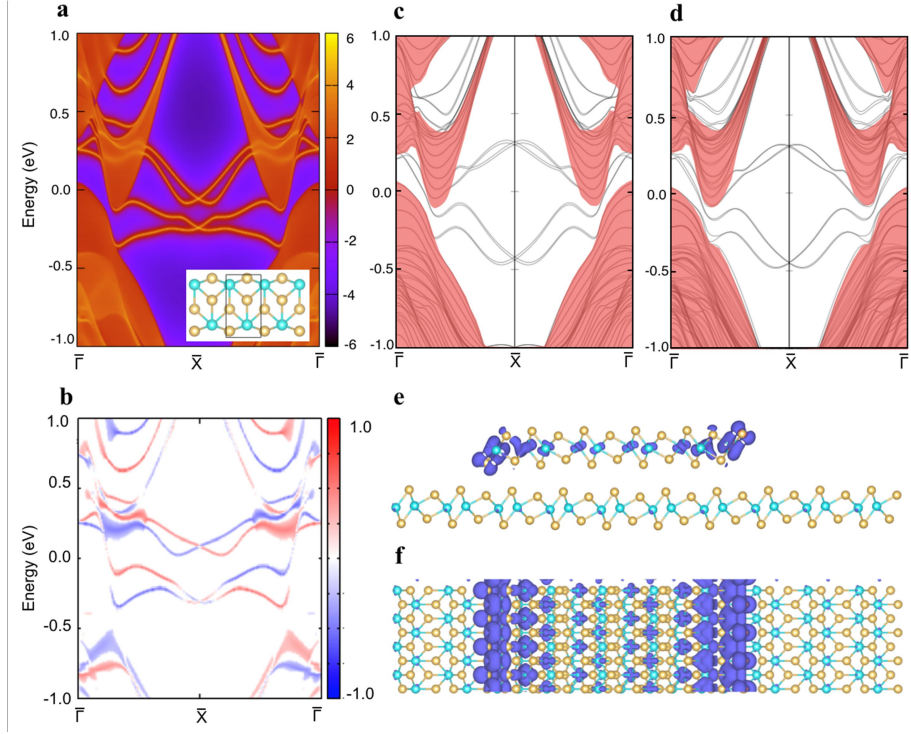

**Figure 4 Calculated topological edge state of WTe$_2$. a,** The topological ES is calculated from the iterative Green's function approach based on a tight-binding model constructed from the monolayer 1T'-WTe$_2$ bulk electronic structure. **b,** The spin polarization of the edge states, here the red and blue color characterize their opposite spin z-components. As also indicated in **a** the edge states in the two Dirac cones with same spin polarization connect to each other at $\bar{\Gamma}$. Thus, they are both topological edge states. **c,** The electronic structure of monolayer 1T'-WTe$_2$ with fully relaxed edges calculated with *ab-initio* method. **d,** Same as **c**, but from a step edge with another bottom layer of 1T'-WTe$_2$ added. The bottom layer is periodic in both *a* and *b* directions. The region colored in light-red in **c** and **d** represents the states from the bulk, whereas the rest are the states staying at the edges. **e,f,** The charge distribution of the ES in real-space from (**e**) the side view and (**f**) the top view.

**References:**

1.  Hasan, M. Z. & Kane, C. L. Topological insulators. *Rev. Mod. Phys*. **82**, 3045-3067 (2010).

2.  Qi, X. L. & Zhang, S. C. Topological insulators and superconductors. *Rev. Mod. Phys*. **83**, 105-1110 (2010).

3.  Kane,C.L.& Mele,E.J. Quantum spin Hall effect in graphene. *Phys. Rev. Lett*. **95**, 226801(2005).

4.  Ren, Y.F., Qiao, Z.H. & Niu, Q. Topological phases in two-dimensional materials: a review. *Rep. Prog. Phys.* **79**, 066501 (2016)

5.  König M. *et al*. Quantum Spin Hall Insulator State in HgTe Quantum Wells, *Science* **318**, 766 (2007).

6.  Knez I., Du R.R., and Sullivan G., Evidence for Helical Edge Modes in Inverted InAs 1=4 Gasb Quantum Wells, *Phys. Rev. Lett*. **107**, 136603 (2011).

7.  Yang F. *et al.* Spatial and energy distribution of topological edge states in single Bi(111) bilayer. *Phys. Rev. Lett.* **109**, 016801 (2012).

8.  Drozdov, I.K. *et al.* One-dimensional topological edge states of bismuth bilayers. *Nat. Phys.* **10**, 664 (2014).

9.  Li X.B.. *et al.*. Experimental observation of topological edge states at the surface step edge of the topological insulator $ZrTe_5$. *Phys. Rev. Lett.* **116**, 176803 (2016).

10. Wu, R. *et al.* Evidence for topological edge states in a large energy gap near the step edges on the surface of ZrTe$_5$. *Phys. Rev. X* **6**, 021017 (2016).

11. C. Pauly *et al.* Subnanometre-Wide Electron Channels Protected by Topology, *Nat. Phys.* **11**, 338 (2015).

12. Qian, X.F., Liu, J.F., Fu L., Li, J. Quantum spin Hall effect in two-dimensional transition metal dichalcogenides. *Science* **346**, 1344-1347 (2014).

13. Zhang, C.X. *et al.* Charge Mediated Reversible Metal–Insulator Transition in Monolayer MoTe$_2$ and W$_x$Mo$_{1-x}$Te$_2$ Alloy. *ACS Nano* **10**, 7370–7375 (2016).

14. Keum, D.H. *et al.* Bandgap opening in few-layered monoclinic MoTe$_2$. *Nat. Phys.* **11**, 482-486 (2015).

15. Lin, Y-C., Dumcenco, D. O., Huang, Y-S. & Suenaga, K. Atomic mechanism of the semiconducting-to-metallic phase transition in single-layered MoS$_2$. *Nat. Nanotech.* **9**, 391–396 (2014).

16. Naylor, C. H. *et al.* Monolayer Single-Crystal 1T'-MoTe$_2$ Grown by Chemical Vapor Deposition Exhibits Weak Antilocalization Effect. *Nano Lett.* 16, 4297-4304 (2016).

17. Chen, J. *et al.* Growth, stabilization and conversion of semi-metallic and semiconducting phases of MoTe$_2$ monolayer by molecular-beam epitaxy. arXiv:1612.06105.

18. Zheng, F.P *et al.* On the Quantum Spin Hall Gap of Monolayer 1T'-WTe$_2$. *Adv. Mater.* **28**, 4845-4851 (2016).

19. Fei, Z.Y. *et al*. Edge conduction in monolayer WTe$_2$. arXiv:1610.07924

20. Soluyanov, A. A. *et al.* Type-II Weyl semimetals. *Nature* **527**, 495-498 (2015).

21. Fu, L. and Kane, C. L. Time Reversal Polarization and a Z$_2$ Adiabatic Spin Pump. *Phys. Rev. B* **74**, 195312 (2006).

22. Fukui, T. and Hatsugai, Y., Quantum Spin Hall Effect in Three Dimensional Materials: Lattice Computation of Z$_2$ Topological Invariants and Its Application to Bi and Sb. *J. Phys. Soc. Jpn.* **76**, 053702 (2007).

23. Soluyanov, A.A. and Vanderbilt, D. Computing topological invariants without inversion symmetry. *Phys. Rev. B* **83**, 235401 (2011)

24. Yu, R., Qi, X.L., Bernevig, A., Fang, Z., Dai, X. Equivalent expression of Z$_2$ topological invariant for band insulators using the non-Abelian Berry connection. *Phys. Rev. B* **84**, 075119 (2011)

25. Kang, D.F. *et al*. Superconductivity emerging from a suppressed large magnetoresistant state in tungsten ditellurid. *Nat. Commun.* **6**, 7804 (2015).

26. Pan, X. C. *et al*. Pressure-driven dome-shaped superconductivity and electronic structural evolution in tungsten ditelluride. *Nat. Comun.* **6**, 7805 (2015).

27. Wang, Z.F. *et al*. Topological edge states in a high-temperature superconductor FeSe/SrTiO$_3$(001) film. *Nat. Mater.* **15**, 968 (2016).

28. Tang, S. *et al*. Realization of Quantum Spin Hall State in Monolayer 1T'-WTe$_2$. arXiv:1703.03151 (2017)

29. Jia, Z.Y. *et al*. Direct Visualization of 2D Topological Insulator in Single-layer 1T'-WTe$_2$. arXiv:1703.04042 (2017)

30. Fu, Y.S., Hanaguri, T., Igarashi, K., Kawamura, M., Bahramy, M.S. & Sasagawa, T. Observation of Zeeman effect in topological surface state with distinct material dependence. *Nat. Commun.* **7**, 10829 (2016).

31. Bloechl, P. E. Projector augmented-wave method. *Phys. Rev. B* **50**, 17953 (1994).

32. Kresse, G. and Joubert, D. From ultrasoft pseudopotentials to the projector augmented wave method. *Phys. Rev. B* **59**, 1758 (1999).

33. Kresse, G. and Hafner, J. Ab initio molecular dynamics for liquid metals. *Phys. Rev. B* **47**, 558 (1993).

34. Kresse, G. and Furthmller, J. Efficiency of ab-initio total energy calculations for metals and semiconductors using a plane-wave basis set. *Comput. Mat. Sci.* **6**, 15 (1996).

35. Perdew, J. P., Burke, K. and Ernzerhof, M. Generalized Gradient Approximation Made Simple. *Phys. Rev. Lett.* **77**, 3865 (1996).

36. Sancho, M.P.L., Sancho, J.M.L. and Rubio, J. Highly convergent schemes for the

calculation of bulk and surface Green functions. *J. Phys. F: Met. Phys.* **15**, 851 (1985).

37. Mostofi, A.A., Yates, J.R., Lee, Y.-S, Souza, I., Vanderbilt, D. and Marzari, N. wannier90: a tool for obtaining maximally-localised Wannier functions. *Comput. Phys. Comm.* **178**, 685 (2008).

38. Franchini, C. *et al.* Maximally localized Wannier functions in LaMnO$_3$ within PBE+U, hybrid functionals and partially self-consistent GW: an efficient route to construct ab initio tight-binding parameters for *eg* perovskites. *J. Phys.: Condens. Matter* **24**, 235602 (2012).

# Supplementary Information

# Observation of topological states residing at step edges of WTe$_2$

Lang Peng, Yuan Yuan, Gang Li[#], Xing Yang, Jing-Jing Xian, You-Guo Shi, Chang-Jiang Yi, Ying-Shuang Fu[*]

Email: [*] yfu@hust.edu.cn   [#] ligang@shanghaitech.edu.cn

## 1. Electronic structure of monolayer 1T'-WTe$_2$

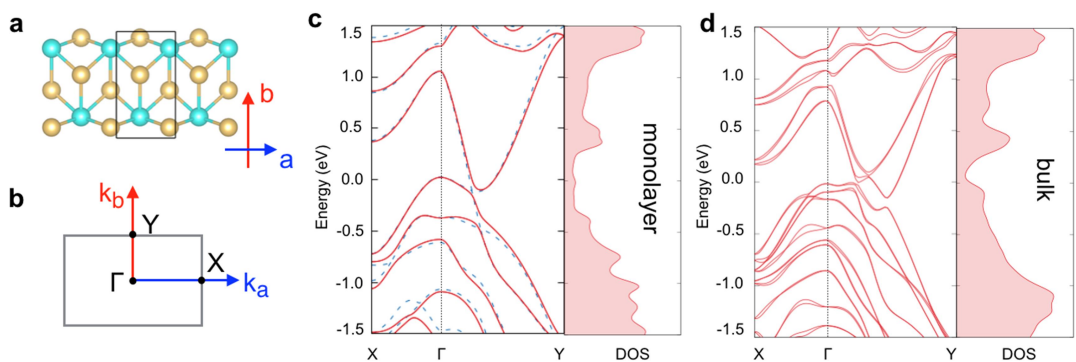

**Figure S1 Electronic structure of 1T'-WTe$_2$ monolayer. a**，The crystal structure and **b** the corresponding Brillouin zone of 1T'-WTe$_2$. **c**, The electronic structure of 1T'-WTe$_2$ (blue dashed line) displays a band crossing slightly below the Fermi level between Γ and Y, which is gapped out by the SOC (solid red line). The corresponding density of states is shown on the right. **d,** Same as **c**, but for the bulk 1T'-WTe$_2$.

The monolayer structure of 1T'-WTe$_2$ shown in Fig. S1a was taken from its bulk crystal structure and was fully relaxed before the electronic structure was calculated. The bulk electronic structure of 1T'-WTe$_2$ monolayer, shown as the blue dashed line in Fig. S1c, features the system a gapless semimetal when the SOC is absent. The top of valence bands and the bottom of conduction bands appear separately at Γ and a k-point between Γ-Y. The inclusion of the SOC gaps out the band crossing residing between Γ and Y, giving rise to the metallic ground state of 1T'-WTe$_2$ monolayer. While, a momentum-dependent Fermi energy can still be found to fully separate the valance from the conduction, which validates the definition of a $Z_2$ topological invariant in this system that will be discussed in supplementary section 3.

2. **Tunneling spectra of edge state along *a* direction at different locations**

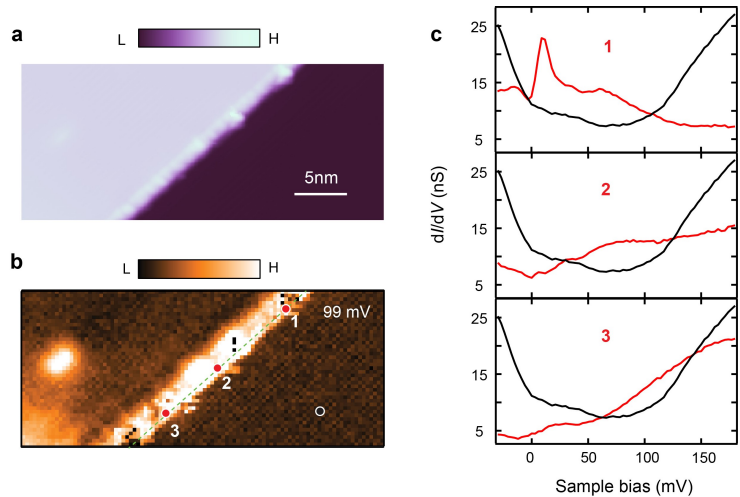

**Figure S2 Spectra of edge state at different locations of *a*-edge. a,** Topography of the single layer high step edge along *a* direction. **b,** Spectroscopic mapping of the step edge in **a** at 99 mV. The data in **a** and **b** are the same as in Fig 2 of main text. **c,** Tunneling spectra (red curves) at different locations of the step edge (red dots in **b**). The spectroscopy (black curve) of the inner terrace (black dot in **b**) is shown for comparison.

# 3. Topological characterization and bulk-edge correspondence.

## 3.1 $Z_2$ topological invariant from topological obstruction and hybrid Wannier charge center

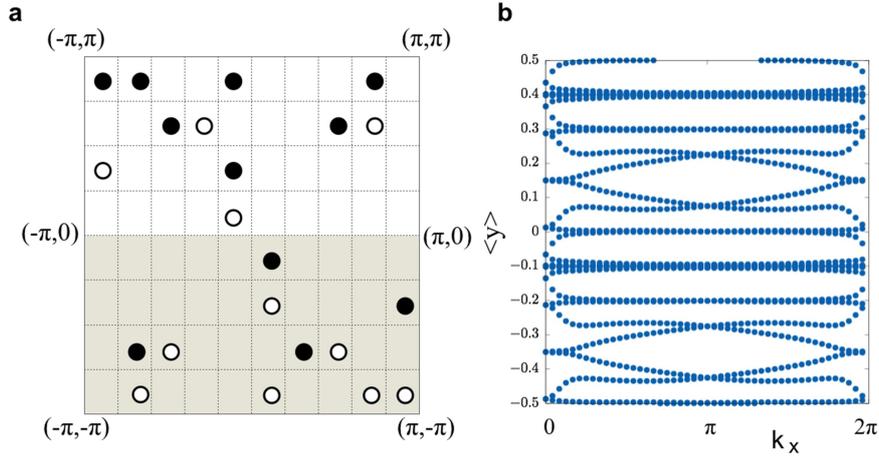

**Figure S3 Topological characterization of monolayer 1T'-WTe$_2$. a**, $Z_2$ topological invariant as an obstruction of smoothly defining the phase of wave function in half of the BZ. **b**, The evolution of the hybrid Wannier charge center <y> as a function of $k_x$.

The clear separation of the conduction and valence bands everywhere in the Brillouin zone (BZ) (Fig. S1c) allows us to define the $Z_2$ topological invariant as usual for this time-reversal symmetric system. Here we apply two different numerical algorithms to characterize the topology of this system, *i.e.* topological obstruction of smoothly defining the phase of the wave function and the hybrid Wannier charge center (Wilson loop).

The band topology of a nontrivial electronic system is defined by the berry curvature of the occupied bands that are separated by an energy gap from the high-energy sector. For topological systems without time-reversal symmetry, the Chern invariant can be viewed as an obstruction to smoothly defining the wave function throughout the entire BZ, which is a torus in 2D. With the time-reversal symmetry present, the time-reversal constraint confines the evaluation of the topological invariant to only half of the BZ, *e.g.* the grey area of the Fig. S3a. It was shown that the topological invariant is given as

$$D = \frac{1}{2\pi}\left[\oint_{\partial\tau} d\ell\, A(k) - \int_{\tau} d\tau\, \mathcal{F}(k)\right] mod\ 2, \qquad (1)$$

where $A(k)$ and $\mathcal{F}(k)$ are the Berry connection and Berry curvature, respectively.

$$A(k) = i\sum_n <u_{k,n}|\nabla_k|u_{k,n}> \text{ and } \mathcal{F}(k) = \nabla_k \times A(k). \qquad (2)$$

$\partial\tau$ and $\tau$ are the corresponding boundary and the area of the grey-color shaded region in Fig. S3a. The above formula can be efficiently evaluated on a discrete lattice consisting of small plaquettes, see Fig. S3a. In each plaquette, Equation (1) takes strictly integer value which is called *n*-filed value, they can be zero (empty), 1 (solid circle) or -1 (empty circle). The $Z_2$ topological invariant is given by (# of solid circle + # of empty circle) modulo 2.

An equivalent way of charactering a $Z_2$ topological insulator is to calculate the hybrid Wannier charge center, as shown in Fig. S3b. The nontrivial connectivity of the Wannier center (symbols), *i.e.* the *n*th curve connects upwards with the (*n+1*)th curve and downwards with the (*n-1*)th curve, characterizes the topology of the system. While the trivial band insulator is qualitatively different from the above picture, where every curve is separated from the others.

## 3.2 Topological edge states of a ribbon along *b*-edge and along *a*-edge with different edge termination

The detailed picture of the ES largely depends on the edge geometry in monolayer 1T'-WTe2, which has a strong implication to the experimental detection of the topological density of states. The topological character, determined by the bulk band structure, is unaffected by the different choices of edge geometry, but the different edge terminations may shift ES in energy, as shown Fig. S4 (along the *b*-edge of the unit cell shown in Fig. S1a) and S5a (along the top and bottom a-edge of the unit cell shown in Fig. S5b). The actual edge termination can be much more complicated than the cases studied in Fig. 1 and Fig. S5b. As a result, the topological ES measured at different spots of the step edge can appear at different binding energies.

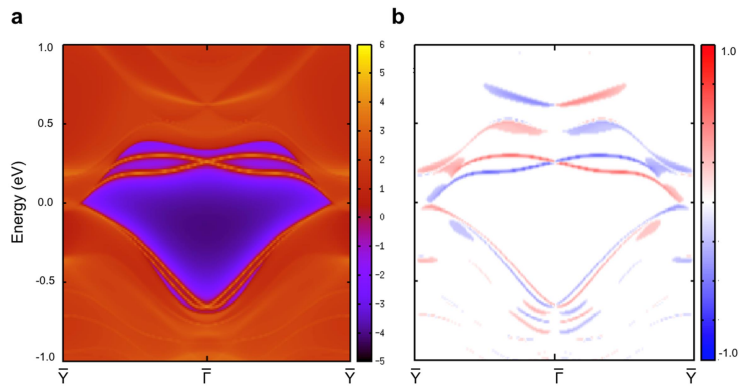

**Figure S4 The topological ES (a) and their corresponding spin z-components (b) of the monolayer 1T'-WTe$_2$ along the *b*-edge of the unit cell shown in Fig. S1a.**

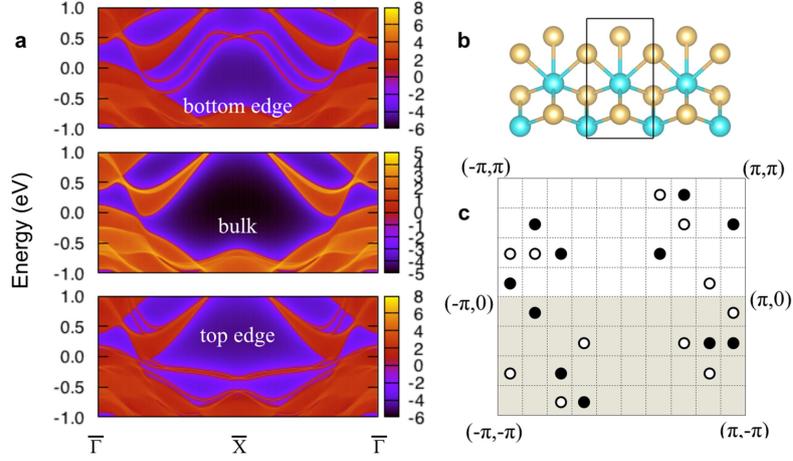

**Figure S5 The ES (a) and the topological character (c) of monolayer 1T'-WTe$_2$ with different edge terminations shown in b.** In **b** the top and bottom edges are of different geometries yielding the different binding-energy locations of the ES. Both ES in **a** are topologically nontrivial, which are proven by the *n*-field configuration in **c** as a topological obstruction to smoothly defining the wave function within half of the 2D BZ (*e.g.* the grey area).

4. Effect of external pressure on the QSH phase

1T'-WTe$_2$ becomes superconducting upon applying external pressure. Lattice constants *a* and *b*, thus the in-plane geometry, are less changed by pressure, however *c* is largely reduced. The emergence of superconductivity demands a significant change of the states around the Fermi level, thus the electronic structure at low energy sector where the nontrivial band topology resides is expected to be modified by pressure. The question as to whether the topology of the monolayer 1T'-WTe$_2$ survives in this case naturally arises.

To construct the monolayer 1T'-WTe$_2$ under pressure, we extracted the lattice constants of the bulk 1T'-WTe2 at 5.36GPa and 10.934GPa from Ref. 28 and fully relaxed the internal coordinates of every atom. From the converged structure, we extracted one layer of WTe$_2$ as the monolayer and added sufficient vacuum in the *c*-direction. We kept the relative position of each atom to be the fully relaxed position of the bulk. Repeating the electronic structure calculation and the topological analysis explained before, we found that, at the two pressures studied, the topological character of the monolayer 1T'-WTe$_2$ is unchanged, *i.e.* the systems remain as QSH semimetals. This conclusion is kept unchanged even after we further relaxed the internal coordinates of each atom in the monolayer. This is expected, as the interlayer coupling of WTe$_2$ is weak and the change of *a* and *b* is negligibly small. In Tab. S1 and Tab. S2, we give the coordinates of each atom in bulk and in monolayer geometry used in our calculations.

| bulk | Monolayer from bulk | Monolayer fully relaxed |
|---|---|---|
| WTe2 5.36GPa bulk<br>1.00<br>  6.157000 0.0000 0.0000<br>  0.000000 3.4060 0.0000<br>  0.000000 0.0000 13.2479<br>    W    Te<br>    4    8<br>Direct<br> 0.89891 0.5000 0.00068<br> 0.10109 0.0000 0.50068<br> 0.54122 0.0000 0.98457<br> 0.45878 0.5000 0.48457<br> 0.29240 0.5000 0.10273<br> 0.70761 0.0000 0.60273<br> 0.79987 0.0000 0.15205<br> 0.20014 0.5000 0.65205<br> 0.35473 0.0000 0.33315<br> 0.64527 0.5000 0.83315<br> 0.85268 0.5000 0.38267<br> 0.14732 0.0000 0.88267 | WTe2 5.36GPa mon. from bulk<br>1.0<br>  6.1570 0.0000 0.0000<br>  0.0000 3.4060 0.0000<br>  0.0000 0.0000 52.992<br>    W    Te<br>    2    4<br>Direct<br> 0.10109 0.0000 0.12517<br> 0.45878 0.5000 0.12114<br> 0.70761 0.0000 0.15068<br> 0.20013 0.5000 0.16301<br> 0.35473 0.0000 0.08329<br> 0.85268 0.5000 0.09567 | WTe2 5.36GPa mon. relaxed<br>1.0000<br>  6.15700 0.00000 0.00000<br>  0.00000 3.40600 0.00000<br>  0.00000 0.00000 52.9920<br>    W    Te<br>    2    4<br>Direct<br> 0.10011 0.00000 0.12512<br> 0.45816 0.50000 0.12121<br> 0.70842 0.00000 0.15191<br> 0.20137 0.50000 0.16339<br> 0.35699 0.00000 0.08292<br> 0.84997 0.50000 0.09441 |

**Table S1: The crystal structure of the bulk and monolayer 1T'-WTe$_2$ at 5.36GPa.**

| bulk | Monolayer from bulk | Monolayer fully relaxed |
|---|---|---|
| WTe2 10.934GPa bulk<br>1.00000<br>  6.12200 0.00000 0.00000<br>  0.00000 3.33500 0.00000<br>  0.00000 0.00000 12.78999<br>  W  Te<br>  4   8<br>Direct<br> 0.88343 0.50000 0.00067<br> 0.11657 0.00000 0.50067<br> 0.52217 0.50000 0.98459<br> 0.47783 0.00000 0.48459<br> 0.27288 0.50000 0.10609<br> 0.72712 0.00000 0.60609<br> 0.78244 0.00000 0.15708<br> 0.21756 0.50000 0.65708<br> 0.36706 0.00000 0.32819<br> 0.63294 0.50000 0.82819<br> 0.86865 0.50000 0.37923<br> 0.13135 0.00000 0.87923 | WTe2 10.934GPa mon. from bulk<br>1.0<br>  6.12200 0.00000 0.00000<br>  0.00000 3.33500 0.00000<br>  0.00000 0.00000 51.15998<br>  W  Te<br>  2   4<br>Direct<br> 0.11657 0.00000 0.12517<br> 0.47783 0.50000 0.12115<br> 0.72711 0.00000 0.15152<br> 0.21756 0.50000 0.16427<br> 0.36706 0.00000 0.08205<br> 0.86865 0.50000 0.09481 | WTe2 10.934GPa mon. relaxed<br>1.0<br>  6.12200 0.00000 0.00000<br>  0.00000 3.33500 0.00000<br>  0.00000 0.00000 51.15998<br>  W  Te<br>  2   4<br>Direct<br> 0.11571 0.00000 0.12507<br> 0.47568 0.50000 0.12124<br> 0.72704 0.00000 0.15358<br> 0.21947 0.50000 0.16515<br> 0.37245 0.00000 0.08117<br> 0.86444 0.50000 0.09275 |

**Table S2: Same as Tab. S1 but at 10.934GPa.**